\documentclass[10pt,prl,showpacs,twocolumn,superscriptaddress]{revtex4}%
\usepackage[dvips]{graphics,color}
\usepackage{amsmath}
\usepackage{amssymb}
\usepackage{graphicx}
\usepackage{amsfonts,times}
\usepackage{amsfonts}%
\begin{document}
\title{Exotic quantum phase transitions in a Bose-Einstein condensate coupled to an
optical cavity}
\author{Gang Chen}
\affiliation{Department of Physics and Center of Theoretical and Computational Physics, The
University of Hong Kong, Pokfulam Road, Hong Kong, China}
\affiliation{Institute of Theoretical Physics, Shanxi University, Taiyuan 030006, China}
\author{Xiaoguang Wang}
\affiliation{Zhejiang Institute of Modern Physics, Department of Physics, Zhejiang
University, Hangzhou 310027, China}
\author{J. -Q. Liang}
\affiliation{Institute of Theoretical Physics, Shanxi University, Taiyuan 030006, China}
\author{Z. D. Wang}
\email{zwang@hkucc.hku.hk}
\affiliation{Department of Physics and Center of Theoretical and Computational Physics, The
University of Hong Kong, Pokfulam Road, Hong Kong, China}

\begin{abstract}
A new extended Dicke model, which includes atom-atom interactions
and a driving classical laser field, is established for a
Bose-Einstein condensate inside an ultrahigh-finesse optical cavity.
A feasible experimental setup with a strong atom-field coupling is
proposed, where most parameters are easily controllable and thus the
predicted second-order superradiant-normal phase transition may be
detected by measuring the ground-state atomic population. More
intriguingly, a novel \textit{second-order} phase transition from
the superradiant phase to the \textquotedblleft Mott" phase is also
revealed. In addition,
a rich and exotic phase diagram is presented.

\end{abstract}

\pacs{03.75.Kk, 42.50.Pq}
\maketitle

As is known, a trapped Bose-Einstein condensate (BEC) may be used to generate
a macroscopic quantum object consisting of many atoms that are in the same
quantum state with a longer lifetime and can be excited by either deforming
the trap or varying the interactions among atoms. Thus the BEC, as a distinct
macroscopic quantum system, plays an important role in the in-depth
exploration of both fundamental physics and quantum device applications of
many-body systems
\cite{1}. In particular, an intriguing idea to combine the cavity quantum
electrodynamics (QED) with the BEC has recently attracted significant
interests both theoretically and experimentally as many exotic quantum
phenomena closely related to both light and matter at ultimate quantum levels
may emerge \cite{2,3,4,5,6,7,8,9,10,11,12,13}.

Very recently, a so-called strong coupling of a BEC to the quantized
field of an ultrahigh-finesse optical cavity was realized
experimentally~\cite{14}, which not only implies that a new
challenging regime of cavity QED has been reached, where all atoms
occupying a single mode of a matter-wave field that can couple
identically to the photon induced by the cavity mode, but also opens
a wider door to explore a variety of new quantum phenomena
associated with the cavity-mediated many-body physics of quantum
gas. Regrettably, the authors of Ref.~\cite{14} ignored the
important nonlinear interactions among the untracold atoms that are
controllable via the Feshbach resonance technique, while these
interactions are believed to have also a considerable impact on
physical properties of the BEC, leading to some exotic quantum
phenomena~\cite{15}.

In this Letter, we establish an extended Dicke model with the
atom-atom interactions and a driving classical laser field under the
two-mode approximation. A feasible experimental setup with
controllable parameters including a collective strong atom-field
coupling is proposed. We illustrate how to drive a well-known
second-order superradiant-normal phase transition
and how to detect it experimentally.
Remarkably, this superradiant phase transition was predicted in quantum optics
many years ago, but has never been observed in experiments
\cite{16,17,18,19,20,21,22}. More intriguingly, a novel
\textit{second-order} superradiant to \textquotedblleft Mott" phase transition
is also revealed. In addition, we also obtain a rich and exotic phase diagram
covering phenomena from quantum optics to the BEC, which is attributed to the
competition between the atom-atom and the atom-field interactions.
\begin{figure}[ptbh]
\includegraphics[width=0.35\textwidth]{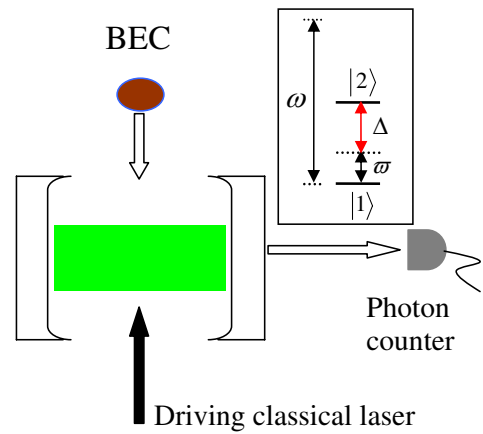}\newline\caption{(Color
online) Schematic diagram of an experimental setup for a BEC of $^{87}$Rb
atoms coupled to a QED cavity. The BEC with two levels $\left\vert
1\right\rangle $ and $\left\vert 2\right\rangle $ is prepared in a
time-averaged, orbiting potential magnetic trap. After moving the BEC into an
ultrahigh-finesse optical cavity, an external controllable classical laser is
applied to produce various transitions of the atoms between $\left\vert
1\right\rangle $ and $\left\vert 2\right\rangle $ states. }%
\label{fig:setup}%
\end{figure}

Our proposed experimental setup is depicted in Fig. 1. For an optical cavity
with length $176$ $\mu$m and the mode waist radius $27$ $\mu$m, we may choose
the parameters of the cavity $(g_{0},\kappa,\gamma)=2\pi\times(10.6,1.3,3)$
MHz \cite{14}, where $g_{0}$ is the maximum single atom-field coupling
strength, $\kappa$ and $\gamma$ are the amplitude decay rates of the excited
state and the intracacity field, respectively. Such a choice implies that the
system is in the strong coupling regime, and thus the long-range coherence
could be well established and the quantum dissipation effect may be safely
neglected. Based on a pair of coupled Gross-Pitaevskii equations for the BEC
with two levels $\left\vert F=1,m_{f}=-1\right\rangle $ $(\left\vert
1\right\rangle )$ and $\left\vert F=2,m_{f}=1\right\rangle $ $(\left\vert
2\right\rangle )$ of $5^{2}S_{1/2}$ \cite{23} and under the two-mode
approximation, the total Hamiltonian for the elastic two-body collisions with
the interaction potential of $\delta$-functional type may be written as
\begin{equation}
\hat{H}=H_{ph}+H_{at-ph}+H_{at}+H_{at-cl}+H_{at-at} \label{1}%
\end{equation}
with $H_{ph}=\omega a^{\dagger}a$ ($\hbar=1$ hereafter), $H_{at}=\omega
_{1}c_{1}^{\dagger}c_{1}+(\omega_{2}+\omega_{12})c_{2}^{\dagger}c_{2}$,
$H_{at-at}=\eta_{1}c_{1}^{\dagger}c_{1}^{\dagger}c_{1}c_{1}/2+\eta_{2}%
c_{2}^{\dagger}c_{2}^{\dagger}c_{2}c_{2}/2+\chi c_{1}^{\dagger}c_{1}%
c_{2}^{\dagger}c_{2}$, $H_{at-cl}=\Omega\lbrack c_{2}^{\dagger}c_{1}%
\exp(-i\varpi t)+c_{1}^{\dagger}c_{2}\exp(i\varpi t)]/2$, and $H_{at-ph}%
=\tilde{\lambda}(c_{1}^{\dagger}c_{2}+c_{2}^{\dagger}c_{1})(a^{\dagger}+a)$,
where $a$ is the annihilation operator of the cavity mode with frequency
$\omega$; $c_{1}$ and $c_{2}$ are the annihilation boson operators for
$\left\vert 1\right\rangle $ and $\left\vert 2\right\rangle $, respectively;
$\omega_{l}=\int d^{3}\mathbf{r}\{\phi_{l}^{\ast}(\mathbf{r})[-\nabla
^{2}/2m_{R}+V(\mathbf{r})]\phi_{l}(\mathbf{r})$ $(l=1,2)$with $V(\mathbf{r})$
being a single magnetic trapped potential of frequencies $\omega_{i}(i=x,y,z)$
and $m_{R}$ being the atomic mass; $\omega_{12}$ is the atomic resonance
frequency; $\eta_{l}=(4\pi\rho_{l}/m_{R})\int d^{3}\mathbf{r}\left\vert
\phi_{l}(\mathbf{r})\right\vert ^{4}$ and $\chi=(4\pi\rho_{1,2}/m_{R})\int
d^{3}\mathbf{r}\left\vert \phi_{1}(\mathbf{r})\right\vert ^{2}\left\vert
\phi_{2}(\mathbf{r})\right\vert ^{2}$ with $\rho_{l}$ and $\rho_{1,2}%
(=\rho_{2,1})$ being the intraspecies and the interspecies $s-$wave scattering
lengths, respectively; $\Omega=2\Omega_{0}\int d^{3}\mathbf{r}\phi_{2}^{\ast
}(\mathbf{r})\phi_{1}(\mathbf{r})$ with $\Omega_{0}$ being the Rabi frequency
for the introduced classical laser with a driving frequency $\varpi$; and
$\tilde{\lambda}=\tilde{g}\int d^{3}\mathbf{r}\phi_{2}^{\ast}(\mathbf{r}%
)\phi_{1}(\mathbf{r})=\tilde{g}\int d^{3}\mathbf{r}\phi_{1}^{\ast}%
(\mathbf{r})\phi_{2}(\mathbf{r})$ with $\tilde{g}$ being a interaction
constant between the atom and the photon \cite{24}.

Under a unitary transformation $U=\exp(-i\varpi J_{z} t)$ with the condition
$\varpi<< \omega$ and using the Schwinger relations $J_{x}=(c_{2}^{\dagger
}c_{1}+c_{1}^{\dagger}c_{2})/2$, $J_{y}=(c_{1}^{\dagger}c_{2}-c_{2}^{\dagger
}c_{1})/2i$, and $J_{z}=(c_{1}^{\dagger}c_{1}-c_{2}^{\dagger}c_{2})/2$ with
the Casimir invariant $J^{2}=N(N/2+1)/2$, Hamiltonian (1) can approximately be
rewritten as%
\begin{equation}
H=\omega a^{\dagger}a+\frac{\lambda}{\sqrt{N}}J_{x}(a^{\dagger}+a)+\omega
_{0}J_{z}+\Omega J_{x}+\frac{v}{N}J_{z}^{2} \label{2}%
\end{equation}
in the rotating frame, where $\lambda=2\tilde{\lambda}\sqrt{N}$ denotes a
collective coupling strength, $v=N[(\eta_{1}+\eta_{2})/2-\chi]$ describes the
atom-atom interactions including the repulsive $(v>0)$ and attractive $(v<0)$
interactions, and $\omega_{0}=\omega_{2}-\omega_{1}+(N-1)(\eta_{2}-\eta
_{1})/2+\Delta$ with $\Delta=\omega_{12}-\varpi$ being the detuning. For a
single trapped potential, we have $\omega_{2}=\omega_{1}$ and consider only
the case of $\rho_{1}=\rho_{2}$, which has the advantages that it reduces the
effects of fluctuations in the total atomic number and ensures a large spatial
overlap of different components of the condensate wavefunction. Thus, the
parameters $v$ and $\omega_{0}$ can further be reduced to $v=N(\eta_{1}-\chi)$
and $\omega_{0}=\Delta$. Eq.~(2) is a key result, which describes the
collective dynamics for the composite system and has a rich phase diagram.
Here we refer this equation to as an \textit{extended Dicke model} since it
contains the extra laser field term (the 4-th one) and atom-atom interaction
term (the 5-th one) in comparison with the standard Dicke model and its
generalized version~\cite{16,20}.

A distinct property of Hamiltonian (2) lies in that all parameters can be
controlled independently. For example, the effective coupling strength
$\lambda$ can be manipulated by a standard technique. The effective Rabi
frequency $\Omega$ and the detuning $\Delta$ depend on the experimentally
controllable classical laser, and especially, the detuning $\Delta$ can vary
continuously from the red $(\Delta<0)$ to the blue $(\Delta>0)$ detunings. The
parameter $v$ ranging from the positive to the negative is determined by the
$s-$wave scattering lengths via Feshbach resonance technique \cite{15}. For
$v=0$ $(\rho_{1,2}=\rho_{1})$ and $\Omega=\varpi=0$, Hamiltonian (2) is
reduced to a standard Dicke model with a second-order superradiant phase
transition at the critical point $\lambda_{c}=\sqrt{\omega\omega_{0}}%
$~\cite{16,17,18,19,20,21,22}. It should be noticed that this important
prediction has never been observed in experiments. The main difficulties are
likely (i) all atoms can hardly interact identically with the same quantum
field; (ii) the frequencies $\omega$ and $\omega_{0}$ typically exceed the
coupling strength $\lambda$ by many orders of magnitude; (iii) it is hard to
control the parameters as demanded. However, in our proposal, these
difficulties could be completely overcome by using the currently available
experimental techniques of BEC, as will readily be seen below.

To explore quantum phases and their transitions, we now investigate the
ground-state properties of Hamiltonian (2), which can approximately be dealt
with by using the Holstein-Primakoff transformation, $J_{\dagger}=b^{\dagger
}\sqrt{N-b^{\dagger}b}$, $J_{-}=\sqrt{N-b^{\dagger}b}b$ and $J_{z}%
=(b^{\dagger}b-N/2)$ with $[b,b^{\dagger}]=1$. Here we introduce two
shifting boson operators $c^{\dagger}=a^{\dagger}+\sqrt{N}\alpha$
and $d^{\dagger }=b^{\dagger}-\sqrt{N}\beta$ with auxiliary
parameters $\alpha$ and $\beta$ to describe the collective behaviors
of both the atoms and the photon \cite{19,20,21,22}. With the help
of the boson expansion method, the scaled ground-state energy is
given by $E_{0}(\alpha,h)/N=\omega\alpha
^{2}-2\lambda\alpha(h^{2}-1/2)+\Delta h\sqrt{1-h^{2}}+\Omega(h^{2}%
-1/2)+vh^{2}(1-h^{2})$ with $h\sqrt{1-h^{2}}=\beta^{2}-1/2$
($1/2\leq h^{2}\leq 1$). The critical points can be determined from
the equilibrium condition $\partial\lbrack
E_{0}(\alpha,h)/N]/\partial \alpha=0$ and $\partial\lbrack
E_{0}(\alpha,h)/N]/\partial h\times dh/d\beta$=0, which leads to two
equations: $\alpha=\lambda(\eta^{2}-1)/2\omega(\eta^{2}+1)$ and
\begin{equation}
2(u+v)\eta(1-\eta^{2})+2\Omega\eta(1+\eta^{2})+\Delta(1-\eta^{4})=0, \label{3}%
\end{equation}
where $\eta=h/\sqrt{1-h^{2}}$ and $\ u=\lambda^{2}/\omega$ are introduced as
new parameters for convenience. The coefficient ($u+v)$ describes the
intrinsic competition between the atom-atom and the atom-field interactions
and gives rise to some exotic phase transitions predicted in the following.

\begin{figure}[ptbh]
\includegraphics[width=0.45\textwidth]{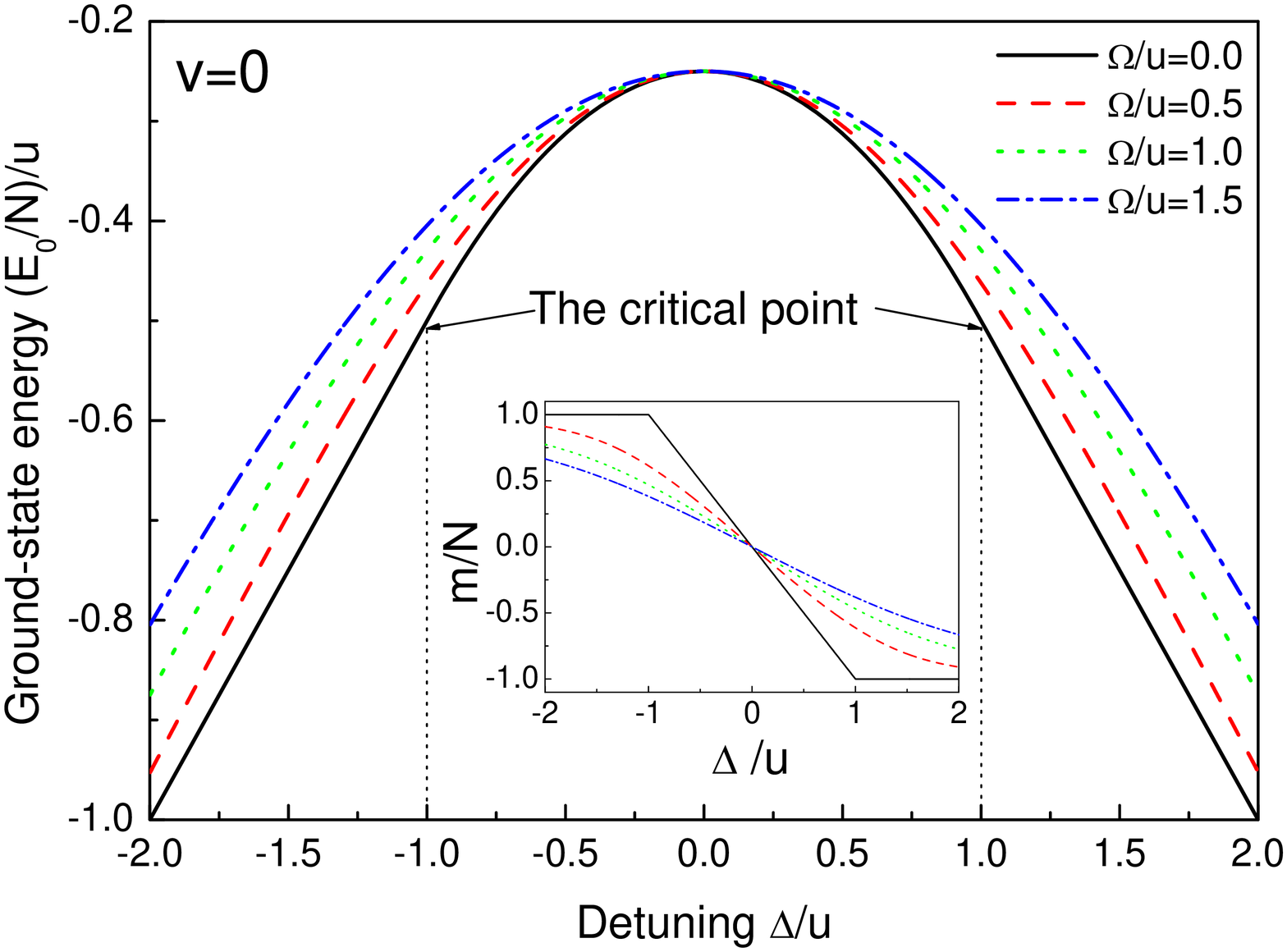}\newline
\caption{(Color online) The scaled ground-state energy $E_{0}/N$ and atomic
population $m/N$ (Insert part) versus the detuning $\Delta$ for different Rabi
frequencies ($\Omega$) at $v=0$. }%
\label{fig:setup}%
\end{figure}

Equation (3) contains the basic information of quantum phases and
their transition. As a benchmark, we first address the simplest case
that there is no nonlinear interaction among atoms, namely, $v=0$
$(\rho_{1}=\rho_{1,2})$. Fig. 2 shows the scaled ground-state energy
$E_{0}/N$ and atomic population (or equivalently \textquotedblleft
magnetization") $m/N$ as a function of the detuning $\Delta$ for
different Rabi frequencies ($\Omega$). It can be seen clearly that
in the limit $\Omega\rightarrow0$, this system exhibits collective
excitations of both the atom and the field with macroscopic
occupations (i.e., $|m/N|<1$ and $\left\langle
a^{\dagger}a\right\rangle >0$) for $-u<\Delta<u$, whereas there are
no such excitations for $\Delta>u$ and $\Delta<-u$ (the solid black
line). This interesting behavior typically shows the second-order
superradiant phase transition in quantum optics with the critical
point $\Delta_{c}=\pm u$ \cite{16,17,18,19,20,21,22}. Moreover, here
we may achieve the condition that the order of magnitude of
$\lambda$ is the same as that of $\sqrt{\omega\Delta}$ by
controlling the detuning of the classical laser. By controlling
 $\Omega/u \ll 1$ and evaluating a partial derivative of $m$ with respect to
 $\Delta$ ( or $\varpi $),  if a peak is detected in the derivative, which becomes sharper and shaper if
 $\Omega$ becomes smaller and smaller, a second order superradiant phase transition at $\Omega =0$ is
 signatured, even though the transition disappears at a
finite Rabi frequency $\Omega$.
 In view of this, our proposed composite system
with the controllable classical laser is a promising candidate for
exploring cavity-induced superradiant phase transition by measuring
the ground-state atomic population via the resonant absorption
imaging~\cite{24}.

\begin{figure}[ptbh]
\includegraphics[width=0.35\textwidth]{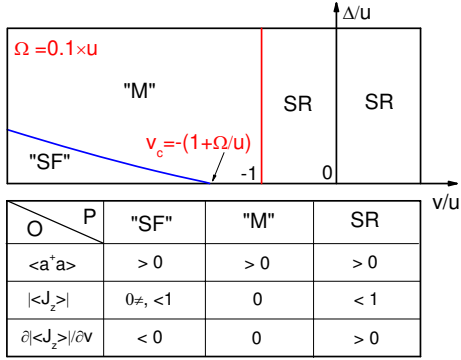}\newline
\caption{(Color online) A zero-temperature phase diagram of the
detuning $\Delta$ and the atom-atom interaction strength $v$ with a
Rabi frequency $\Omega$. The blue line is determined by
$\Delta=\pm\lbrack(u+v)^{2/3}-\Omega^{2/3}]^{3/2}$. Two interesting
second-order phase transitions occur when crossing the red and blue
lines. Note that this phase diagram is symmetric with respect to
$\Delta <0$, simply because the Hamiltonian (2) is invariant under
the transformation $\Delta =-\Delta$ and $J_{z}=-J_{z}$. The lower
Table denotes three different quantum phases.
Here, \textquotedblleft SF"=\textquotedblleft Superfluid" phase,
\textquotedblleft M"=\textquotedblleft Mott" phase,
SR=Superradiant phase, P=phase, and O=order parameter.}%
\label{fig:setup}%
\end{figure}
On the other hand, the nonlinear interactions among atoms controlled
by Feshbach resonance technique play an important role for the
ground-state properties. Fig.3 plots a zero-temperature phase
diagram for the atom-atom interaction strength $v$ and the detuning
$\Delta$ with a Rabi frequency in the framework of mean field. The
Table lists the corresponding ranges of the mean intracavity photon
number $\left\langle a^{\dagger}a\right\rangle $, the atomic
population $\left\langle J_{z}\right\rangle $, and the
\textquotedblleft susceptibility" $\partial\left\langle
J_{z}\right\rangle /\partial v$ for three different quantum phases.
In the case of the repulsive interaction $(v>0)$, the critical point
becomes $\Delta_{c}=\pm(u+v)$, which implies that an effective
atom-field interaction is enhanced, while in the weak attractive
interaction case $(-u<v<0)$, the effective interaction is
suppressed. However, the basic features of the superradiant phases
remain. In particular, in the case of $v=-u$, this system exhibits a
novel \textit{second-order }phase transition from the superradiant
to the \textquotedblleft Mott" phases (Red line) \cite{25}. The
relevant physics can be intuitively understood as following. In an
optical cavity with $\omega>>\lambda$, the cavity mode is only
weakly or virtually excited, and the energy term $\omega
a^{\dagger}a+(\lambda/\sqrt{N})J_{x}(a^{\dagger}+a)$ is therefore
nearly equal to $-uJ_{x}^{2}/N$. If $v>-u$, the ground-state
properties are governed by the energy $-\left\vert u+v\right\vert J_{x}%
^{2}/N+\Delta J_{z}+\Omega J_{x}$. The effective potential in the
Landau-Ginzburg theory is a double-well potential with the photon-assisted
Josephson tunneling, which means that this system is located at the
superradiant phase. If $-u-\Omega<v<-u$, the energy $-\left\vert
u+v\right\vert J_{z}^{2}/N+\Delta J_{z}+\Omega J_{x}$ is dominant and the
corresponding effective potential is a single-well potential with no internal
Josephson tunneling, leading to the same atomic numbers for the two levels
$(m=0)$, which may be referred to as the \textquotedblleft Mott" phase
\cite{26}. Also, when $v$ is decreased, a second-order phase transition from
the \textquotedblleft Mott" to the \textquotedblleft superfluid" phases (Blue
line) occurs at the critical point $v_{c}=-u-[\Omega^{2/3}+\Delta^{2/3}%
]^{3/2}$. In the so-called \textquotedblleft superfluid" case, the effective
potential is another double-well potential with the internal Josephson
tunneling induced by the attractive interaction \cite{26}. It should be
pointed out that these three different phases can be distinguished
experimentally by measuring the atomic population $\left\langle J_{z}%
\right\rangle $ and the \textquotedblleft susceptibility" $\partial
\left\langle J_{z}\right\rangle /\partial v$. In the limit $\Omega
\rightarrow0$, this predicted second-order phase transition from the
superradiant to the \textquotedblleft Mott" phases becomes a direct transition
from the superradiant to the \textquotedblleft superfluid" phases with the
same order at the critical point $v_{c}=-u$ and $\Delta=0$.

Although the second-order superradiant phase transition disappears in the
strong attractive interaction $(v<-u)$, another interesting phase transition
(from the phase with nonzero macroscopic occupation of the level 1 to that of
the level 2) in the \textquotedblleft superfluid" regime emerges when the
detuning $\Delta$ changes from negative to positive (i.e., from the red to the
blue detunings). Fig. 4 shows the scaled atomic population $m/N$ versus
$\Delta$ for different $\Omega$s. We see that a novel \textit{first-order}
\textquotedblleft superfluid" phase transition occurs at $\Delta=0$, and
moreover this first-order phase transition exists until $\Omega_{c}=\left\vert
u+v\right\vert $ (Red dashed line). For $\Omega=$ $\Omega_{c}$, it becomes a
\textit{second-order} phase transition with the same critical point. For
$\Omega>$ $\Omega_{c}$, no phase transition has been seen by varying $\Delta$.

\begin{figure}[ptbh]
\includegraphics[width=0.45\textwidth]{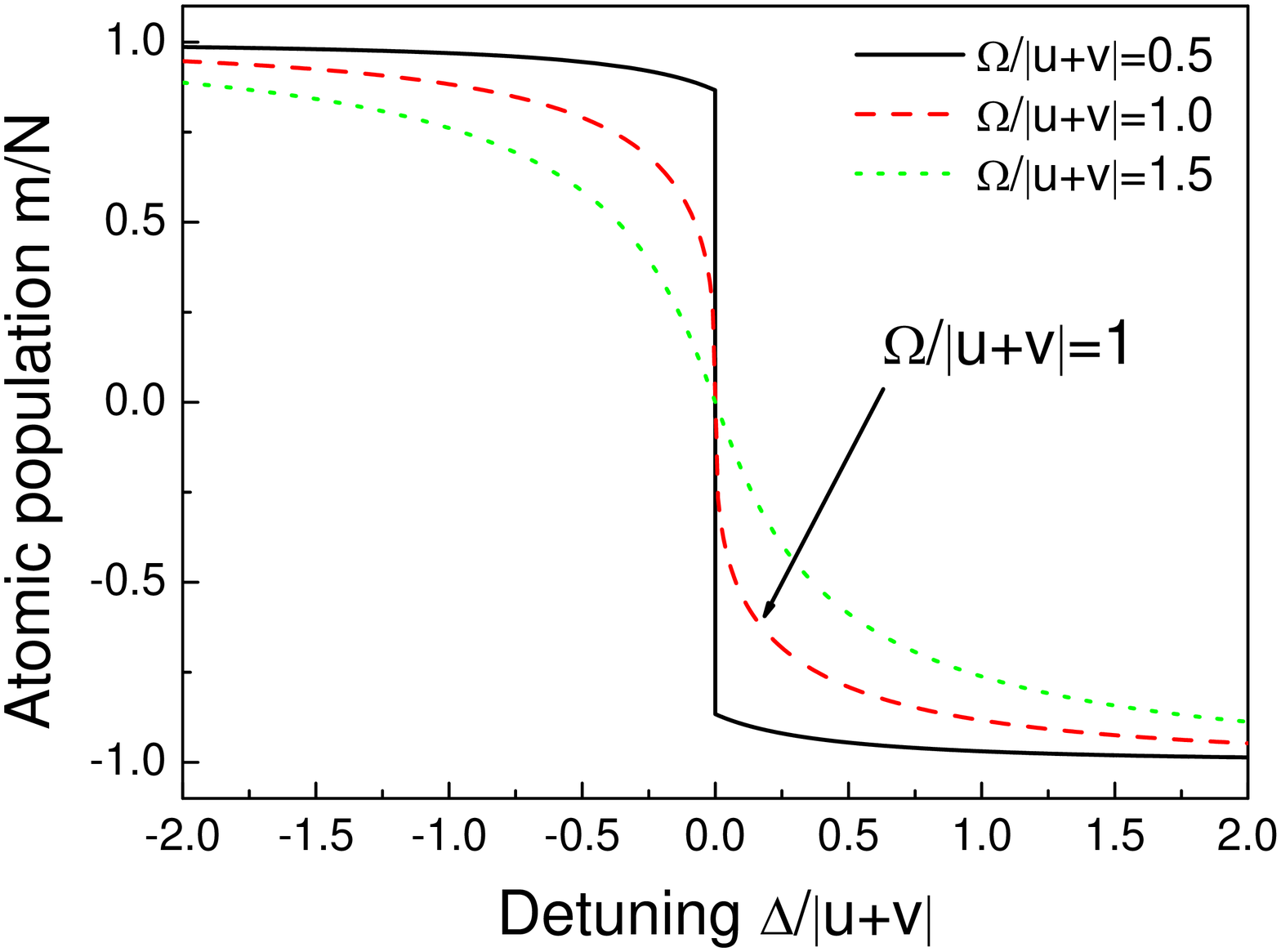}\newline
\caption{(Color online) The scaled ground-state atomic population $m/N$ versus
the detuning $\Delta$ for different Rabi frequencies $\Omega$ when $v<-u$.}%
\end{figure}

We now estimate the energy scales for the parameters in Hamiltonian (2) to
address the experimental feasibility.
Under the two-mode approximation,
the wavefunctions of the macroscopic condensate states for the single magnetic
trap may roughly be approximated by $\phi_{l}(\mathbf{r})=\pi^{-3/4}%
(d_{x}d_{y}d_{z})^{-1/2}\exp[-(x^{2}/d_{x}^{2}+y^{2}/d_{y}^{2}+z^{2}/d_{z}%
^{2})/2]$ with $d_{x}=\sqrt{1/m_{R}\omega_{x}}$, $d_{y}=\sqrt{1/m_{R}%
\omega_{y}}$ and $d_{z}=\sqrt{1/m_{R}\omega_{z}}$. Hence, the atom-atom
interaction strength can be estimated by $v=N(\rho_{1}-\rho_{1,2})/\sqrt{2\pi
}d_{x}d_{y}d_{z}m_{R}$. For the typical values $(\omega_{x},\omega_{y}%
,\omega_{z})=2\pi\times(290,43,277)$ Hz, $\rho_{1}=4.2$ nm, $\rho_{1,2}=9.7$
nm, and $m_{R}=1.45\times10^{-25}$ kg, the energy scale of $v$ is about
$-0.238$ MHz with $N=5\times10^{4}$, which ensures that the error (the order
of $1/\sqrt{N}$) for determining the ground-state properties by means of the
Holstein-Primakoff transformation is very low. The effective coupling strength
$\lambda=2\tilde{\lambda}\sqrt{N}=2.81\times10^{4}$ MHz for $\tilde{\lambda
}=2\pi\times10$ MHz \cite{14} is indeed in the strong coupling regime. The
energy scale for $u$ is about $0.315\ $MHz for $\omega=2.51\times10^{9}$ MHz
\cite{14}, which can be adjusted by controlling the frequency of photon. These
energy scales for $u$ and $v$ imply that the intrinsic competition between the
atom-atom and atom-field interaction should be taken into account seriously in
the BEC coupled to the optical cavity. Also note that the aforementioned
condition $\varpi<< \omega$ is well satisfied once $\varpi$ is tuned around
$\omega_{12}$ since $\omega_{12}\sim6.8 \times10^{3}$ MHz $<< \omega$
\cite{23}.

Finally, we elaborate briefly how to probe the predicted phase transitions
experimentally. From the condition $\alpha=\lambda(\eta^{2}-1)/2\omega
(\eta^{2}+1)$ with $\lambda=2.81\times10^{4}$ MHz and $\omega=2.51\times
10^{9}$ MHz, we can immediately evaluate the maximum of the scaled mean
intracavity photon number $\left\langle a^{\dagger}a\right\rangle /N$ and find
it to be much less than the critical intracavity photon number $n_{c}%
=\gamma^{2}/2g_{0}^{2}=0.04$. Therefore, one is able to perform the
transmission spectroscopy measurement with a weak probe laser to obtain the
ground-state energy spectrum and atomic population since different quantum
phases are, in general, characterized by their specific dispersion relations.
The transmission (of this probe laser through the cavity) versus the detuning
may be monitored and/or detected by counting photons out of the cavity. Only
when the probe laser frequency matches a system in resonance, the
corresponding transmission is anticipated \cite{27}.

In summery, we have established an extended Dicke model and designed
a feasible experimental setup with controllable parameters. An
exotic phase diagram has been obtained, which covers various
phenomena from quantum optics to the BEC and reveals particularly
several novel quantum phase transitions.

We thank S. L. Zhu, Y. Li, D. L. Zhou, L. B. Shao, and Z. Y. Xue for helpful
discussions. This work was supported by the RGC of Hong Kong under Grant No.
HKU7051/06P, the URC fund of HKU, the NSFC under Grant Nos. 10429401, 10775091
and 10704049, and the State Key Program for Basic Research of China (No. 2006CB921800).


\begin{thebibliography}{99}
\bibitem {25}The first-order derivative of the ground-state energy with
respect to $v$ is continuous at $v=-u$, but its second-order derivative has a
discontinuty at the same point.
\end{thebibliography}
\end{document}